\documentclass[ prd,twocolumn]{revtex4}
\usepackage{xcolor}
\usepackage[utf8]{inputenc}
\usepackage{graphicx}
\usepackage{empheq}
\usepackage{braket}
\usepackage{float}
\usepackage{amssymb}
\usepackage{amsmath}
\usepackage{mhchem}
\usepackage{siunitx}
\usepackage[titletoc]{appendix}

\usepackage{hyperref}
\hypersetup{colorlinks=true, linkcolor=red, citecolor=red, urlcolor=purple}

\newcommand\purple[1]{{\color{purple}#1}}

\begin{document}

\title{\purple{Vortex matter in a two-band SQUID-shaped superconducting ﬁlm}}

\author{C. A. Aguirre$^{1,2,}$\footnote[2]{\href{cristian@fisica.ufmt.br}{cristian@fisica.ufmt.br}}\href{https://orcid.org/0000-0001-8064-6351}{\includegraphics[scale=0.05]{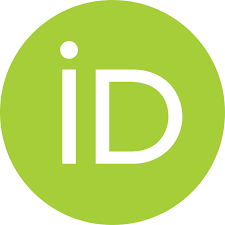}}}

\author{Julián Faúndez$^{2}$\href{https://orcid.org/0000-0002-6909-0417}{\includegraphics[scale=0.05]{Figures/orcid.png}}, S. G. Magalhães$^{2}$\href{https://orcid.org/0000-0002-6874-7579}{\includegraphics[scale=0.05]{Figures/orcid.png}}}
\author{J. Barba-Ortega$^{3,4}$\href{https://orcid.org/0000-0003-3415-1811}{\includegraphics[scale=0.05]{Figures/orcid.png}}}
\affiliation{$^{1}$Departamento de Física, Universidade Federal de Mato-Grosso, Cuiabá, Brasil}
\affiliation{$^{2}$Condensed Matter Physics Group, Instituto de Fisica, Universidade Federal do Rio Grande do Sul, $91501-970$ Porto Alegre, RS, Brazil.}

\affiliation{$^{3}$Departamento de Física, Universidad Nacional de Colombia, Bogotá, Colombia}
\affiliation{$^{4}$Foundation of Researchers in Science and Technology of Materials, Bucaramanga, Colombia}

\date{\today}
\begin{abstract} 
%******************************************************************************
In the present work we studied the magnetization, vorticity, Cooper pairs density and the space distribution of the local magnetic field in a three-dimensional superconductor with a SQUID geometry (a square with a central hole connected to the outside vacuum through a very thin slit). Our investigation was carried out in both the Meissner-Oschenfeld and the Abrikosov state solving the two-band Ginzburg-Landau equations considering a Josephson coupling between the bands. We found a non-monotonic vortex behavior and the respective generation of vortex cluster due to the Josephson coupling used between condensates.
\end{abstract}
%******************************************************************************
\maketitle
%******************************************************************************
\section{Introduction}
%******************************************************************************
The application of superconducting technology in different nano and mesoscopic systems has been of great importance in recent years \cite{1,2}. Special interest has been the low-temperature superconductors (LTS) which have been used in various applications; among them, the production of magnets with high magnetic fields in the range of $5-10$ $T$ \cite{3}, which are used in particle accelerators and in magnetic resonance devices  \cite{4}. Furthermore, LTS's are used in the manufacture of high precision superconducting electronic devices such as microwaves, detectors and superconducting quantum interference devices (SQUID), etc \cite{Fillis-Tsirakis, Clarke}. Due to an increase in the application of different devices superconductors, such devices are expected to show an increase in critical temperatures, the so-called high critical temperature superconductors (HTCS). These implementations range from the application to the management and treatment of extensive databases, to solutions for mobility in high-speed trains around the world. However, the most widely used magnetometer is the SQUID, since it allows measuring even very slight magnetic fields.\\\\
%******************************************************************************
This electronic device is composed of a superconducting plate with a cavity inside and one or more superconductor-insulator-superconductor separators. When the insulating barrier is thin enough between two superconductors there is the possibility that current flows from one superconductor to another, even without the need to apply voltage between them. It was thus that Josephson predicted a super-current proportional to the sine of the phase difference of the order parameters of each superconductor and the consequent dependence of the super-current on the phase differences combined with the presence of magnetic fields makes these instruments highly accurate \cite{Kirtley,Jaklevic,Zimmerman,Jaklevic1}.
%******************************************************************************
A SQUID can be made from a high critical temperature superconductor, such as cuprates \ce{YBa2Cu3O7}, which has a critical temperature  $T_ {c} \approx 92 K$ \cite{Malozemof}, alloys $Nb/Al-AlO_x$,  $Bi_{1-x}Sb_x2Se_3$;  topological insulator $(TI)$ nano-ribbon (NR) connected with $Pb_{0.5}In_{0.5}$ superconducting electrodes \cite{Zhang,Kim,Tsuei,Kirtley1,Tsuei1,Tsuei2}. Additionally, an attempt has been made to adapt and apply this multi-band system to other problems with various techniques that have allowed to study theoretically the vortex state. For example, Rogeri \textit{et al.}, using a genuinely three-dimensional approach to the time-dependent Ginzburg – Landau theory, studied the local magnetic ﬁeld profile of a mesoscopic superconductor in the so-called SQUID geometry. They studied the magnetic induction in both the Meissner and the mixed state as temperature function \cite{Rogeri}. Brandt \textit{et al.}, using the London theory calculated dynamic electromagnetic properties in thin flat superconducting films of rectangular and circular films without and with slits and holes. The sheet currents and the coupling between the vortices and the defects were expressed by a stream function. They found that to the long-ranging magnetic stray field, the interaction energy between vortices and the magnetic field depend on the size, cross section of the sample and shape of the film \cite{Brandt1,Brandt2,ClemBrandt}.\\\\
%*****************************************************************************
The influence of the boundary conditions on the magnetization curve of the sample in a thin mesoscopic superconductor in the SQUID geometry (circular with a hole at the center connected to the outer rim by a very thin slit) was studied in the reference \cite{BarbaSquid}, they found that  the first vortex penetration field and vorticity strongly depend on the boundary condition. H. J. M. ter Brake \textit{et al.}, made an interesting road-map that describes in a general way the developments of superconducting digital electronics under simulations and circuit design, circuit manufacturing and new devices and materials \cite{Brake}. T. Noh  \textit{et al.}, described a SQUID in which the Josephson junctions are formed from strips of metal normal in contact with a superconductor. They measure the flux dependence of the critical current of this system without applying a finite voltage bias across the Superconductor-metal normal-superconductor junction, enabling sensitive flux detection without generating microwave radiation \cite{Noh}. M. Mori \textit{et al.}, studied the $\pi-$SQUID comprising $0-$ and $\pi-$ Josephson junctions, they found that the $\pi-$SQUID can be a $\pi-$Qubit with spontaneous loop currents by which the half-integer Shapiro-Steps are induced, then the $0-$ and $\pi-$Josephson junctions equivalent is a key for the half-integer Shapiro-Steps and realizing the $\pi-$Qubit \cite{Mori}.
%******************************************************************************
Thus, it is difficult to reiterate the extensive importance of this measurement instrument. However, we propose the extension in the manufacture of the same device, but in a two-band extension considering a Josephson type coupling between them. This extension generates a non-monotonic behavior between the vortices and the creation of vortex clusters.\\\\
%******************************************************************************
This article is organized as follows: the theoretical formalism is presented in section \ref{Section1}. In section \ref{Section2} we present the main results for the studied system. We show the vortex states, vorticity, magnetization and profile of the magnetic induction as functions of  external magnetic field $\mathbf{H}$ for a mono and two-band three-dimensional SQUID. Finally, in section \ref{Section3} we detail the main results. 
%**************************************************************************
\section{Theoretical Formalism}\label{Section1}
%******************************************************************************
In this work, we studied the vortex matter in a mesoscopic superconductor in the so-called SQUID geometry, through the functional of a two-band superconductor system. We will consider the interaction between the two bands (or condensates $\psi_{1},\psi_{2}$) in a Josephson type coupling. Thus, the Gibbs energy density for the the superconducting order parameter complex pseudo-function $\psi_{i}=|\psi_{i}|e^{i\theta_{i}}$ ($\theta_{i}$ its phase) \cite{Aguirre,Aguirre1,Egora,Egor2,Egor3}, and magnetic potential $\mathbf{A}$, where $\mathbf{B}=\nabla \times \mathbf{A}$, is:
%******************************************************************************
\begin{eqnarray}
\mathcal{G} = \int dV ( \sum_{i}^{2} \mathcal{F}(\psi_{i},\textbf{A})
+\frac{1}{2 \mu_0} |\nabla \times \mathbf{A}|^2+\Theta(\psi_{i}))\label{Gibbs1}
\end{eqnarray}
%******************************************************************************
where:
%******************************************************************************
{\small
\begin{equation}
\mathcal{F}(\psi_{i},\textbf{A})= \alpha_i |\psi_i|^2 +\frac{\beta_i}{2} |\psi_i|^4+\frac{\zeta_{i}}{2 m_i} |(i \hbar \nabla+2 e \mathbf{A}) \psi_i |^2
\label{Function}
\end{equation}}
%******************************************************************************
and
%******************************************************************************
\begin{eqnarray}
\Theta(\psi_{1},\psi_{2})=\gamma (\psi_1^*\psi_2+\psi_2^*\psi_1)
\label{Jops}
\end{eqnarray}
%******************************************************************************
$\alpha_i=\alpha_{i0}(1-T/T_{ci})$ and $\beta_i$ are two phenomenological parameters, $i=1,2$ in the equations \ref{Gibbs1} and \ref{Function}. We used the Josephson coupling showed in the equation \ref{Jops}. In the London gauge $\nabla \cdot \mathbf{A}=0$. We express the temperature $T$ in units of the critical temperature $T_{c1}$, length in units of the coherence length $\xi_{10}=\hbar/\sqrt{-2 m_1 \alpha_{10}}$, the order parameters in units of $\psi_{i0}=\sqrt{-\alpha_{i0}/\beta_i}$, time in units of the Ginzburg-Landau characteristic time $t_{GL}=\pi\hbar /8k_BT_{c1}$, and the vector potential $\mathbf{A}$ is scaled by $H_{c2}\xi_{10}$, where $H_{c2}$ is the bulk upper critical field. The general form of time dependent Ginzburg-Landau equations for a two-band system in dimensionless units \cite{Egor2,Egor3} is given by:
%***************************************************************************
\begin{align}
\frac{\partial \psi_{1}}{\partial t}=(1-T-|\psi_1|^2) \psi_1-|\mathbf{D}|^2 \psi_1 +\hat{\gamma_{1}}
\label{GL2B1}
\end{align}
%******************************************************************************
\begin{eqnarray}
\frac{\partial \psi_{2}}{\partial t}=(1-\frac{T}{T_{r2}}-|\psi_2|^2)\psi_2- 
\frac{m_{r2}}{\alpha_{r2}} |\mathbf{D}|^2 \psi_2 +\hat{\gamma_{2}}
\label{GL2B2}
\end{eqnarray}
%******************************************************************************
with:
%************************************************************************
\begin{eqnarray}
  \label{GL4}
 \hat{\gamma_{1}}=\frac{\gamma|\psi_{2}|\psi_{1}}{|\psi_{1}|}\bigg[\cos(\theta_{2}-\theta_{1})+ i\sin(\theta_{2}-\theta_{1})\bigg]
 \label{Gamma1}
\end{eqnarray}
%************************************************************************
And
%************************************************************************
\begin{eqnarray}
  \label{GL4}
 \hat{\gamma_{2}}=
\frac{\gamma|\psi_{1}|\psi_{2}}{|\psi_{2}|}\bigg[\cos(\theta_{2}-\theta_{1})-i\sin(\theta_{2}-\theta_{1})\bigg]
\label{Gamma2}
\end{eqnarray}
%************************************************************************
For more details for the  calculus of the relation \ref{Gamma1} and the relation \ref{Gamma2} see  Ref. \cite{Aguirre1}. The equations (4) and (5) are solved in $\Omega_{sc}$. While, the equation for the vector potential $\mathbf{A}$:
%**********************************************************************************
\begin{eqnarray}
\frac{\partial{\bf A}}{\partial t} &=&
\left \{ 
\begin{array}{ll}
{\bf J}_s-\kappa^2\nabla\times\nabla\times {\bf A}\;\;\;\;{\rm in}\;\Omega_{sc}\; \\
-\kappa^2\nabla\times\nabla\times {\bf A}\;\;\;\;{\rm in}\;\Omega\backslash\Omega_{sc}\;
\end{array}
\right .\; \label{EQ5}
\end{eqnarray}
%*****************************************************************************************
are solved in $\partial\Omega_{sc}$, where:
%**************************************************************************************
\begin{equation}
{\bf J}_{s} = \zeta_{1}\Re \left[\psi_1\mathbf{D}\psi_1^*  \right]+ 
\zeta_{2} \Re \left[\frac{\beta_{r2}}{\alpha_{r2}}\psi_2\mathbf{D}\psi_2^*\right]
\label{EQ6}
\end{equation}
%*************************************************************************************
For this case, $\zeta_{1}=\zeta_{2}$, also $\mathbf{D}=i\nabla -\mathbf{A}$. The domain $\Omega_{sc}$ is filled by the superconducting parallelepiped of high $c$ and square lateral sizes $a$ and $b$. The superconducting vacuum interface is denoted by $\partial\Omega_{sc}$. Due the demagnetization effects, we consider a larger domain $\Omega$ of dimensions $A\times B \times C$, such that $\Omega_{sc}\subset\Omega$. The vacuum-vacuum interface is indicated by $\partial\Omega$. The domain $\Omega$ is taken sufficiently large such that the local magnetic field equals the applied field ${\bf H}$ at the surface $\partial\Omega$ or at $x= \pm A /2$, $y= \pm B/2$, and $z= \pm C/2$ planes.
%*************************************************************************************
We have studied a mesoscopic superconducting parallelepiped in the domain $\Omega_{sc}$ of height $c = 1\xi$ and  lateral dimension $a=b=12\xi$; the central hole has dimensions $2.4\xi \times 2.4\xi$. In order to solve the 3D Ginzburg-Landau equations, the size of the simulation box $\Omega$ was taken $C=11\xi$, $A=B=19\xi$. The grid space used was $\Delta_x=\Delta_y=\Delta_z=0.25\xi$ (see reference \cite{sardellaPLA} for more details).\\\\
%*********************************************************************************************************
$\hat{\gamma}_{1},\hat{\gamma}_{2} $ represents the Josephson coupling between the $i$ and $j$ band. The boundary conditions ${\bf n}\cdot(i\mbox{\boldmath$\nabla$}+{\bf A})\psi_{i}=0$, $i=1,2$ with ${\bf n}$ a surface normal outer vector. Also, we defined $m_{r2}=m_{2}/m_{1}=0.5$, $\beta_{r2}=\beta_{2}/\beta_{1}=0.7$, $\gamma=-0.01$, $\zeta_{1}=\zeta_{2}=0.01$, $\kappa=1.0$ \cite{Aguirre,Aguirre1}.
%******************************************************************************
 We choose the zero-scalar potential gauge at all times and use the link variables method for to solve the Ginzburg-Landau equations \cite{Yuriy,Misko2003,Canfield,Time,Time2} (and references therein).
%******************************************************************************
\begin{figure}[htbp!]
\centering
\includegraphics[scale=0.62]{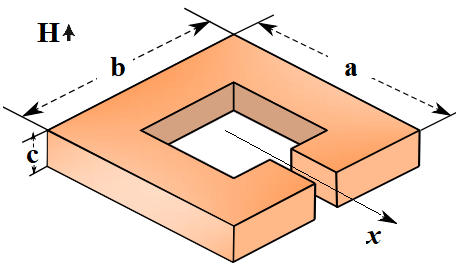}
\caption{Layout of the studied sample in a perpendicular external applied magnetic field $\mathbf{H}$. $c = 1\xi$, $a=b=12\xi$; the central hole has dimensions $2.4\xi \times 2.4\xi$.} 
\label{Sample}
\end{figure}
%******************************************************************************
\section{numeric results}\label{Section2}
%******************************************************************************
In the Fig. (\ref{Sample}) we present the layout of the studied sample. A Squid in a perpendicular external applied magnetic field $\mathbf{H}$ with dimensions $c = 1\xi$, $a=b=12\xi$; the central hole has dimensions $2.4\xi \times 2.4\xi$.  We show the magnetization $-4\pi\mathbf{M}/H_{c2}$, vorticity $N$, superconducting electronic density for the band 1 $|\psi_1|^2$, and for the band 2  $|\psi_2|^2$ respectively and the profile of the magnetic induction $\mathbf{B}$ in $(x,b/2,c/2)$  plane as function of the external applied magnetic field $\mathbf{H}$ when is increasing and decreasing. This loop in the magnetic field, aims to describe a hysteresis cycle and computationally the magnetic field will vary between $ 0\leq \mathbf{H}\leq \mathbf{H}_2$, ($\mathbf{H}_2$ is the upper magnetic field, where we found that $\mathbf{H}_2=2.0$ for the single-band sample and $\mathbf{H}_2=1.8$ for the two-band sample). Finally for magnetization we will use $\mathbf{M}=\mathbf{H} -\mathbf{B}/4\pi $ and for the vorticity or vortex number we use $N=Im(\psi_{i}\nabla \psi_{i}^{*})/(\psi_{i}^{*}\psi_ {i})$, with $i=1,2$, indicating the band index. 
%******************************************************************************
\subsection{Single-band SQUID}
\begin{figure}[htbp!]
\centering
\includegraphics[scale=0.24]{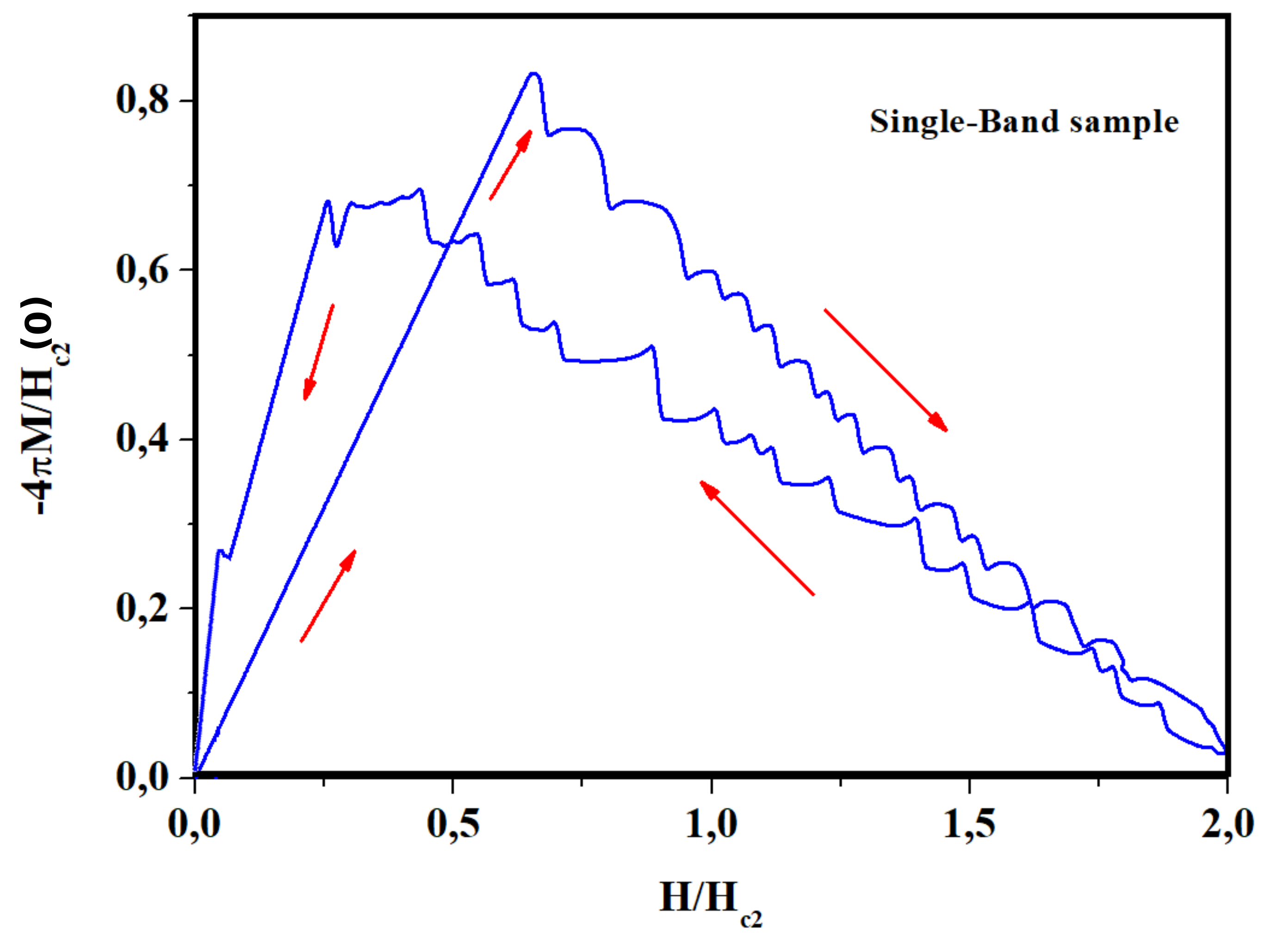}
\caption{Magnetization $-4\pi\mathbf{M}/H_{c2}$ as  function of the external applied magnetic field $\mathbf{H}$ in the upward branch and downward branch for a single-band Squid.}
\label{magmB}
\end{figure}

\begin{figure}[htbp!]
\centering
\includegraphics[scale=0.24]{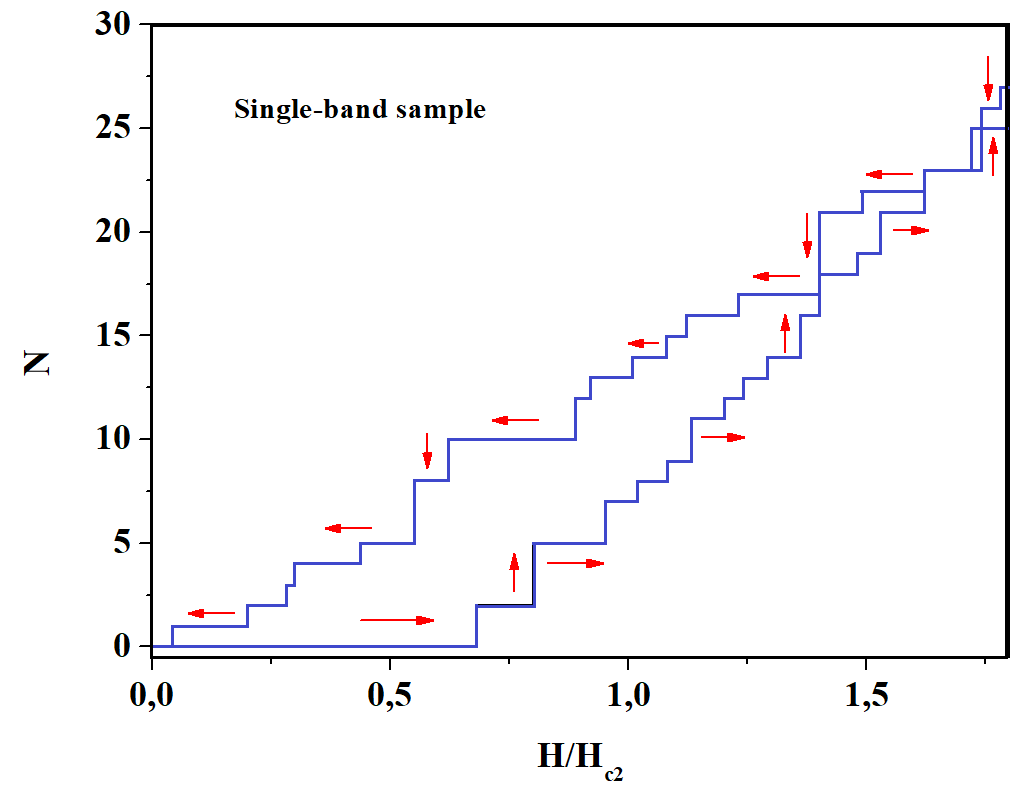}
\caption{Vorticity $N$ (or vortex number) as a function of $\mathbf{H}$ in the upward branch and downward branch for a single-band Squid.}
\label{vnormal}
\end{figure}

\begin{figure*}[htbp!]
\includegraphics[scale=0.3]{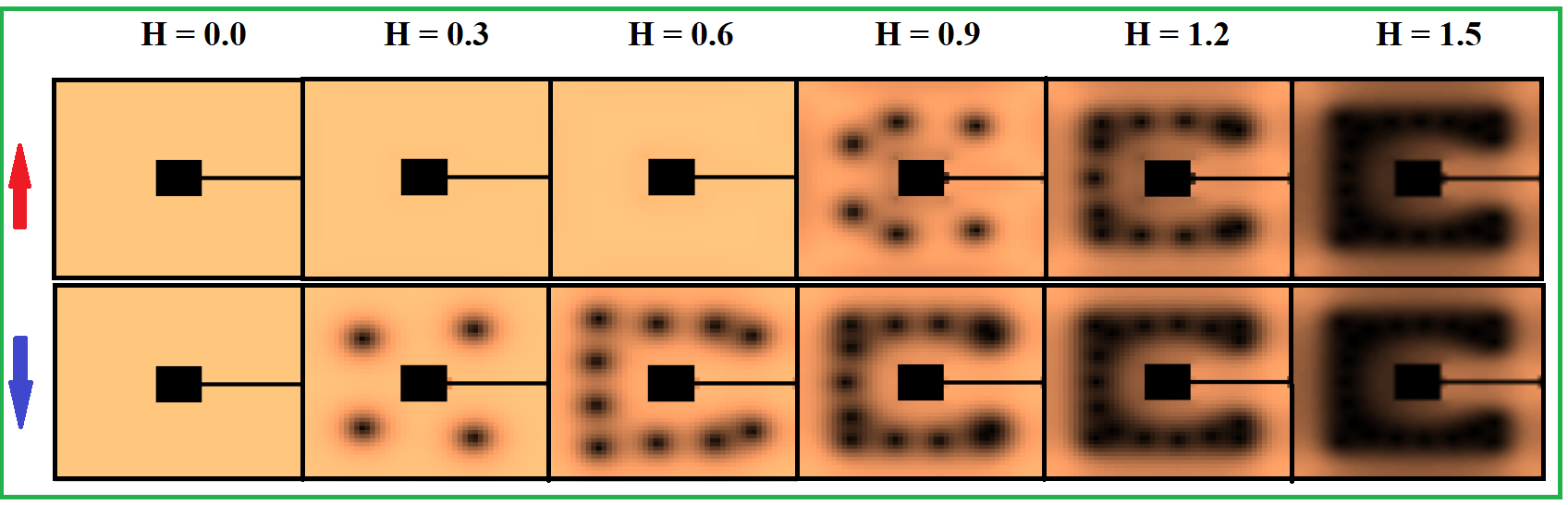}
\caption{Square modulus of the order parameter $|\psi|^{2}$ (density of Cooper pairs/vortex state) as function of the external applied magnetic field $\mathbf{H}$ in the upward branch (upper panels) and downward branch (lower panels), for the indicated values of $\mathbf{H}$ for a single-band Squid sample.} 
\label{normal}
\end{figure*}

%******************************************************************************
In the Fig. (\ref{magmB}) we present the magnetization $-4\pi\mathbf{M}/H_{c2}$, for the single-band Squid system in a loop of the magnetic field. We observe that for the upward branch of the magnetic field, the behavior of the magnetization is conventional. On the other hand in the downward branch of the loop, we observe that the sample does return to the starting point. Additional in $0<\mathbf{H}<\mathbf{H}_1$, ($\mathbf{H}_1\approx 0.7$ is the lower critical field), the sample is in Meissner-Oschenfeld state in the upward branch of $\mathbf{H}$ and $0<\mathbf{H}<0.12$ in the downward branch of $\mathbf{H}$. This behavior has already been studied and in general terms, it is that of a conventional superconducting sample.
%******************************************************************************
%******************************************************************************
In the Fig. (\ref{vnormal}) we present the vorticity $N$ as function of the applied magnetic field $\mathbf{H}$. We observe the quantized entrance of the magnetic flux in the sample, when having a pinning the sample, we observe that in the upward branch, the sample presents different amount of vortices, accounting for the anchoring of vortices by the barrier of energy between the sample and the pinning. Additionally, note that the jump in magnetization in the Fig. (\ref{magmB}), coincides with the in-flux of fluxoids shown in the Fig. (\ref{vnormal}).
%******************************************************************************
%******************************************************************************
Now, in the Fig. (\ref{normal}) we show the vortex state  $(|\psi|^{2})$ (or Density of Cooper pairs), for the single-band sample as function of $\mathbf{H}$ in the upward branch (upper panels) and downward branch (lower panels). Thus, by increasing $\mathbf{H}$ we observe the entry of the vortices in the superconducting sample; now, due to the entry of these vortices there is a competition between the surface energy barrier at the external boundary and the internal pinning, this competition causes the vortices to move in the sample until reaching different positions where the energy is minimized and stabilized in the sample (observe the symmetrical positions with respect to a horizontal mirror $\hat{\sigma}_{h}$ when the field increases and decreases for the same $\mathbf{H}$). This process is repeated with a greater number of vortices as the external magnetic field $\mathbf{H}$ increases, we observe that as the field increases, the loss of the superconducting state is evident (conventional behavior).
%******************************************************************************
%******************************************************************************
\subsection{Two-band SQUID}
\begin{figure}[H]
\centering
\includegraphics[scale=0.235]{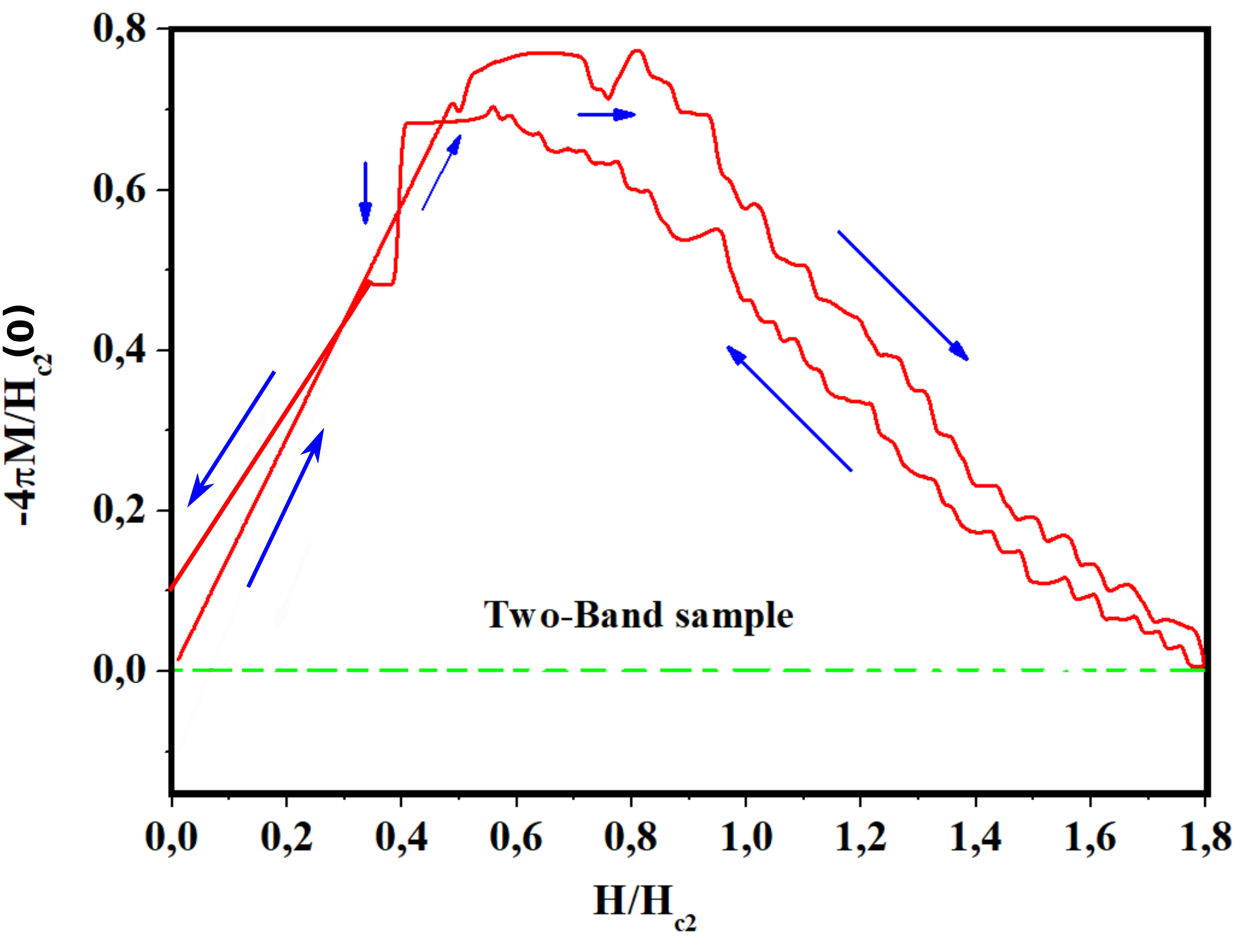}
\caption{Magnetization $-4\pi\mathbf{M}/H_{c2}$ as function of $\mathbf{H}$ in the upward branch and downward branch for a two-band Squid  system.}
\label{M2B}
\end{figure}

\begin{figure}[H]
\centering
\includegraphics[scale=0.235]{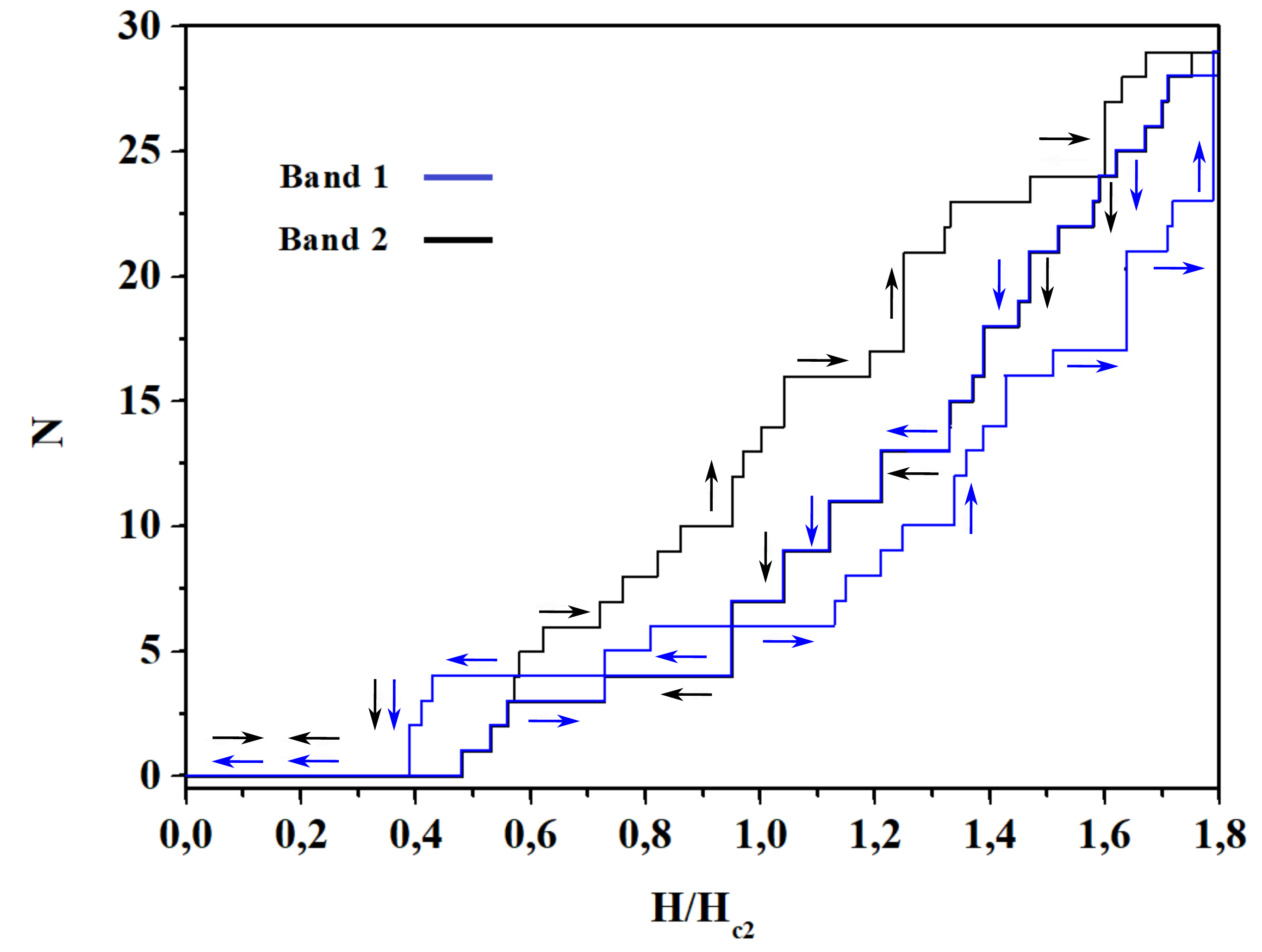}
\caption{Vorticity $N$ as a function of $\mathbf{H}$ in the upward branch and downward branch for a two-band Squid superconductor system.}
\label{V2B}
\end{figure}

\begin{widetext}
\begin{figure*}
\includegraphics[scale=0.11]{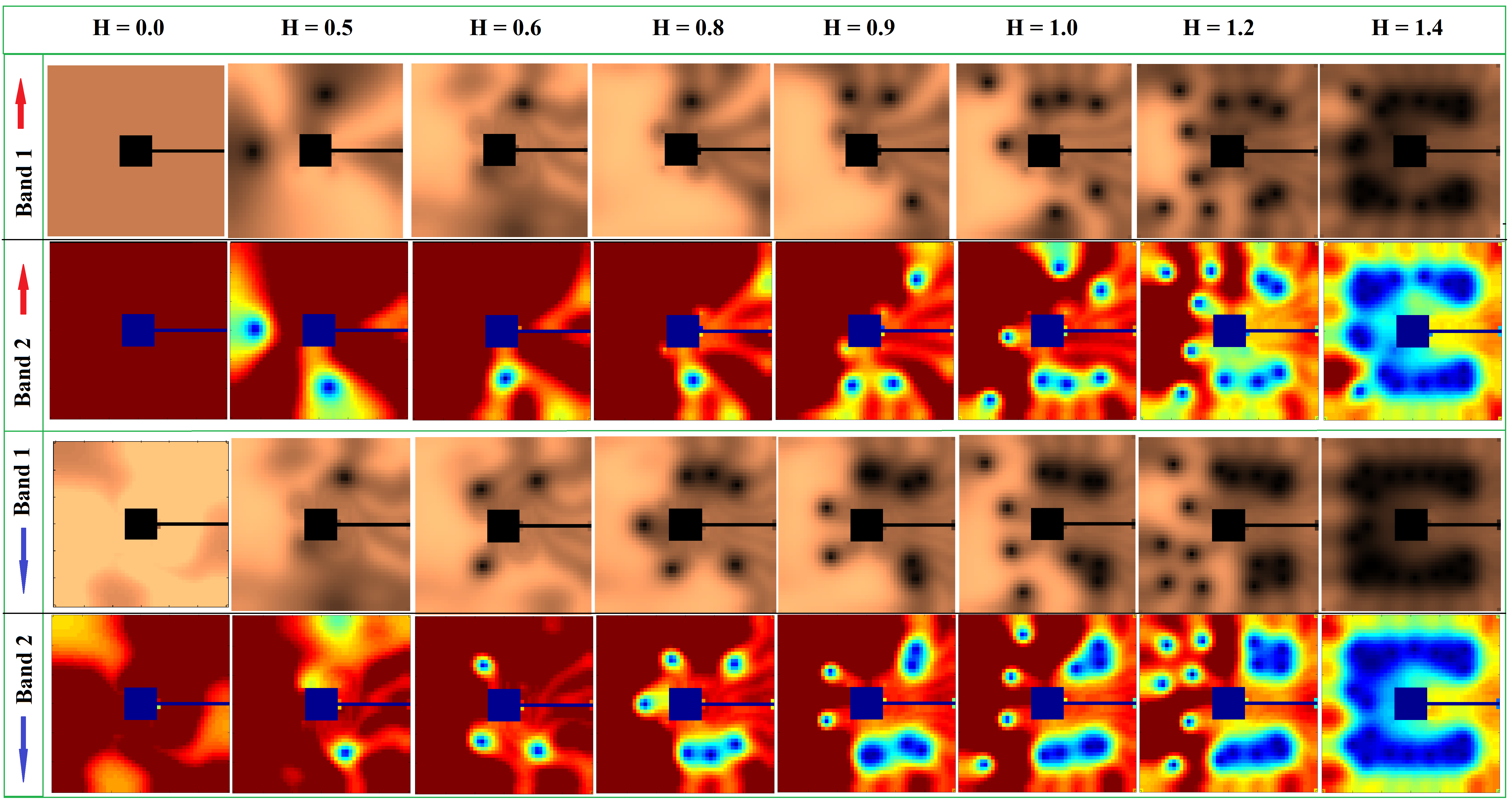}
\caption{Square modulus of the order parameter $|\psi_{1}|^{2},|\psi_{2}|^{2}$ for the band 1 and the band 2 respectively as a function of $\mathbf{H}$ in the upward branch (arrow up) and downward branch (arrow down) for the indicated values of $\mathbf{H}$.} 
\label{vs2b}
\end{figure*}
\end{widetext}

\begin{widetext}
\begin{figure*}
\includegraphics[scale=0.46]{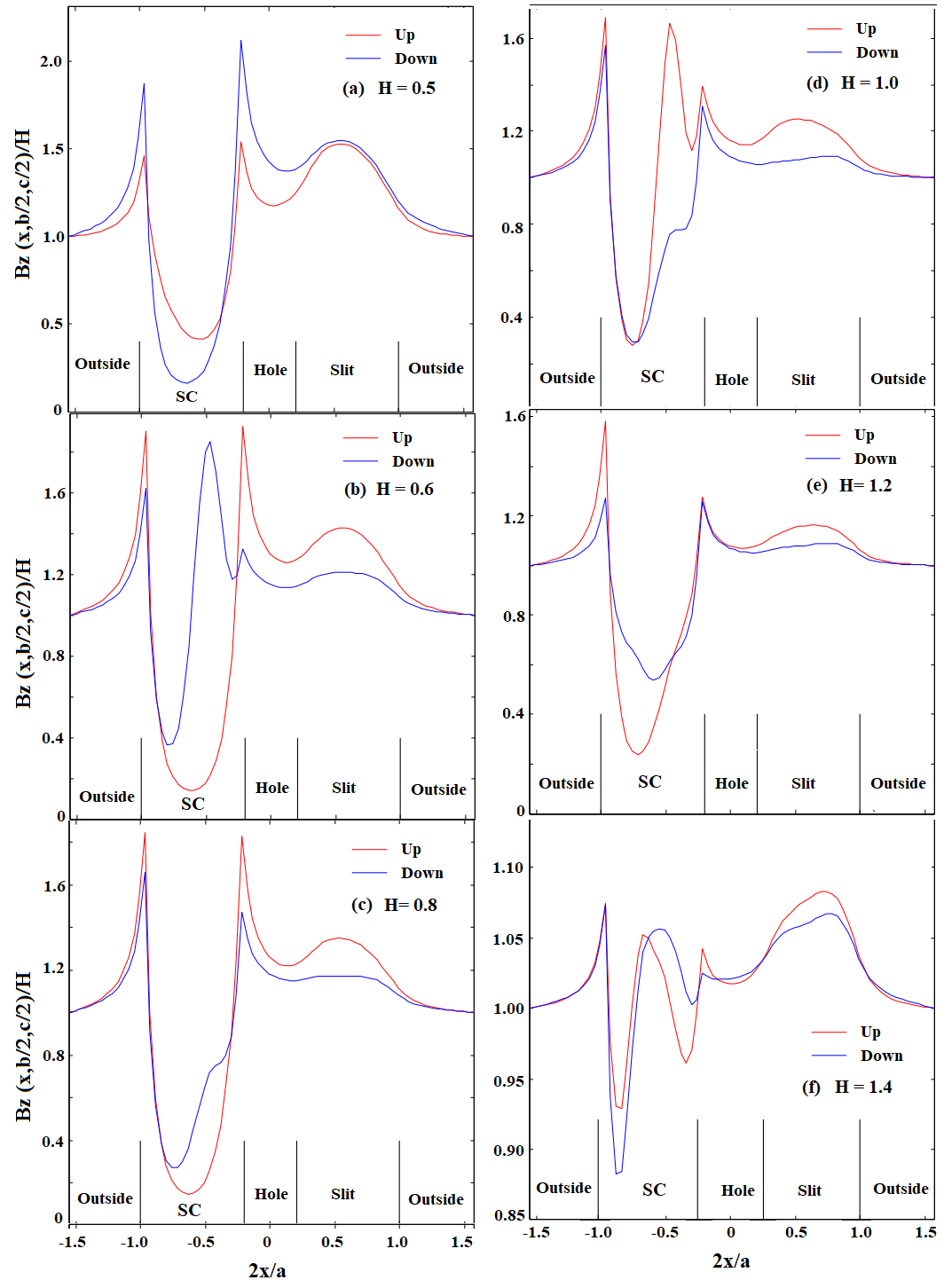}
\caption{Intensity of the $z$ component of the magnetic induction $\mathbf{B}$, normalized to $\mathbf{H}$, along the axis $x$ for indicates values of $\mathbf{H}$ in the upward branch (up) and downward branch (down) for a two-band Squid. (SC means Superconductor).} 
\label{Perfil}
\end{figure*}
\end{widetext}

%******************************************************************************
Now we concentrate on the study of a superconductor SQUID composed of two-bands $\psi_{1},\psi_{2}$ (or two condensates). To do this, we will start at Fig (\ref{M2B}), where we present the magnetization $-4\pi\mathbf{M}/H_{c2}$ curve. Initially we note that for $0<H<H_1$, ($H_1\approx 0.5$ is the lower critical field), it is in the Meissner-Oschenfeld state and that for $ 0.5<\mathbf{H}<0.8$ there is an non-conventional behavior in the system, having drops in magnetization and a successive increase; initially, we observe that the lower critical fields are different for the single-band and two-band systems; $\mathbf{H}_{1-Single-band}>\mathbf{H}_{1-Two-band}$ and  $\mathbf{H}_{2-Single-band}>\mathbf{H}_{2-Two-band}$. After this, the system behaves in a similar way to single-band Squid, in the sense that there are jumps in magnetization, which are indicators of the entry of fluxoids into the sample. In the downward branch of $\mathbf{H}$ (lower panels) we have an non-conventional behavior, even with very strong variations in the magnetization drops. However, for $-4\pi\mathbf{M}/H_{c2}$, we observe the effect of non-homogeneity in the sample, observing the anchoring of vortices in the superconducting sample, an effect due to the competition between the energy barriers in the present system.
%******************************************************************************
%******************************************************************************
Now in the Fig. (\ref{V2B}), we present the vorticity $N$ as function of $\mathbf{H}$ for the bands 1 and 2. Initially we observe that the vorticity $N$ is different for each of the condensates, with which we expect a different vortex states. Now, as the sample is an overlay of the superconducting condensates, the vortex centers are not coincident, giving possible fractional states in the total vortex state (result of the two-band Josephson coupling type), which allows tunneling of vortices and anti-vortices between superconducting bands. Of greater importance is the Fig. (\ref{V2B}), where we observe a collective behavior in phase of the two superconducting bands; this means that when the two-band system passes to a normal state at $\mathbf{H}_{2}=2.0$, the phases and the superconducting condensates stabilize and generate a behavior in phase, establishing a stable and organized loss of the vortices; as you vary in the upward branch for $\mathbf{H}$, this is a novel behavior in a two-band Squid system, where a decrease in vorticity would be expected with different values for each of the condensates.
%******************************************************************************
%******************************************************************************
Now, in the Fig. (\ref{vs2b}) we present the square modulus of the order parameter $|\psi_{1}|^{2},|\psi_{2}|^{2}$ or Cooper pairs density for each band. We can see from the Fig. (\ref{vs2b}) that at $\mathbf{H}=0$, the sample remains in the Meissner-Oschenfeld state, when the magnetic field increases, the vortices enter through the border of the regions closest to the slit, then they move entering the sample until they reach the normal state at $\mathbf{H}_{c2}$. Then, when the magnetic field decreases, the vortices are expelled from the sample until $\mathbf{H}=0$ where we found $N=0$, so, there are not any pinning effect due the presence of the hole. In addition to producing hysteresis in the relaxation of the electronic states present in the superconductor SQUID system.
%******************************************************************************
%******************************************************************************
Now, in the the Fig. (\ref{Perfil}) we have plotted the local magnetic field or magnetic induction $\mathbf{B}$ along the $x$ axis, which passes through the slit and the hole of the SQUID as is shown in the  Fig. (\ref{Perfil}), for the indicates values of $\mathbf{H}$ in the upward branch and downward branch for a two-band case. As can be seen from this graph, the profile of the magnetic
field $B_z(x, b/2, c/2)/H$ is very  dependent on $\mathbf{H}$ in both upward branch and downward branch. We found that in the point in which is calculated the profile $(x, b/2, c/2)$, the magnetic induction profile is identical for both bands, that is, the condensates act in phase under the application of an external magnetic field $\mathbf{H}$, even though the magnetic field in each condensate presents different values and different vortex state (See Fig (\ref{vs2b})).
%******************************************************************************
%********************************************
\section{Conclusions}\label{Section3}
%************************************
In the present work we have studied the magnetic properties and vortex state in a superconducting three dimensional SQUID single and two-band considering a Josephson-type coupling. This study was carry out solving the Ginzburg-Landau time-dependent equations. We study magnetization, vorticity, vortex state and profile of the magnetic induction, observing  non-conventional behavior of the vortex state in the two-band system. Additionally, we present a the typical behavior of the vortex matter for a single-band sample but a non-conventional vortex state for a two-band case, which establish the existence of non-monotonic interaction between the vortices (short-range repulsion and long-range attraction). Finally we not found any difference in the magnetic profile of the magnetic induction between the bands. Our results in the mono-band case are in  qualitative nature agreement with previous works.
%********************************************
\section*{ACKNOWLEDGMENTS}\label{Section4}
%********************************************
C. A. Aguirre, would like to thank the Brazilian agency CAPES, for financial support, Grant number: 0.89.229.701-89. J. Faúndez and S. G. Magalhães thank FAPERGS, CAPES and CNPq for partially financing this work under the Grant PRONEX 16/0490-0. 
%******************************************************************************

\end{document}